\documentclass[3p]{elsarticle}

\usepackage{color,colortbl}
\usepackage{amsmath}
\usepackage{array}
\usepackage{multirow}
\usepackage{moreverb}
\usepackage{graphicx}

\RequirePackage{subfigure}[1995/03/06]

\usepackage[dvips,colorlinks,bookmarksopen,bookmarksnumbered,citecolor=red,urlcolor=red]{hyperref}
\usepackage{color}
\usepackage{todonotes}

\usepackage{amssymb}

\biboptions{longnamesfirst,comma}

\journal{Computational Statistics and Data Analysis}

\begin{document}

\begin{frontmatter}



\title{A Dynamic Bi-orthogonal Field Equation Approach for Efficient Bayesian Calibration of Large-Scale Systems\tnoteref{t1}}

\tnotetext[t1]{This work was supported by Basic Science Research Program through the National Research Foundation of Korea (NRF) funded by the Ministry of Education, Science and Technology (2010-0025484).
}

\author[kaist]{Piyush M.~Tagade\fnref{fn1}}
\ead{piyush.tagade@kaist.ac.kr}

\author[kaist]{Han-Lim~Choi\corref{cor1}\fnref{fn2}}
\ead{hanlimc@kaist.ac.kr}

\cortext[cor1]{Corresponding Author}
\fntext[fn1]{Postdoctoral Research Fellow}
\fntext[fn2]{Assistant Professor}

\address[kaist]{Division of Aerospace Engineering, KAIST, Daejeon 305-701, Republic of Korea}

\begin{abstract}
This paper proposes a novel computationally efficient dynamic bi-orthogonality based approach for calibration of a computer simulator with high dimensional parametric and model structure uncertainty.
The proposed method is based on a decomposition of the solution into mean and a random field using a generic Karhunnen-Loeve expansion.
The random field is represented as a convolution of separable Hilbert spaces in stochastic and spatial dimensions that are spectrally represented using respective orthogonal bases.
In particular, the present paper investigates generalized Polynomial Chaos bases for stochastic dimension and eigenfunction bases for spatial dimension.
Dynamic orthogonality is used to derive closed form equations for the time evolution of mean, spatial and the stochastic fields.
The resultant system of equations consists of a partial differential equation (PDE) that define dynamic evolution of the mean, a set of PDEs to define the time evolution of eigenfunction bases, while a set of ordinary differential equations (ODEs) define dynamics of the stochastic field. This system of dynamic evolution equations efficiently propagates the prior parametric uncertainty to the system response. The resulting bi-orthogonal expansion of the system response is used to reformulate the Bayesian inference for efficient exploration of the posterior distribution.
Efficacy of the proposed method is investigated for calibration of a 2D transient diffusion simulator with uncertain source location and diffusivity.
Computational efficiency of the method is demonstrated against a Monte Carlo method and a generalized Polynomial Chaos approach.
\end{abstract}


\end{frontmatter}


\section{Introduction}
Recent advancements in digital technologies have facilitated the use of computer simulators for investigation of large scale systems.
However, computer simulators are fraught with uncertainties due to poorly known/unknown model, parameters, initial and boundary conditions etc. \cite{OreskesScience94}.
Various researchers have investigated effect of these uncertainties on the credibility of a computer simulator and established uncertainty quantification and calibration as an integral aspect of a modeling and simulation process \cite{Mehta91,Mehta96,OberkampfRESS02,TrucanoRESS06}.
This paper focuses on the Bayesian approach that provide a formal framework to identify, characterize and quantify the uncertainties, and provides a generic inference method for calibration of a computer simulator using limited and noisy experimental data \cite{KennedyJRSS01,HigdonJSC04,Goldstein04,Bayarri_Tech07,KellyRESS09}.

The Bayesian framework is preferred over more traditional calibration methods due to its ability to provide complete posterior statistics of the parameters of interest.
Sampling techniques such as Markov Chain Monte Carlo (MCMC) method \cite{BesagSS95,Gamerman} are used for exploration of the posterior statistics, especially for calibration of nonlinear dynamical simulators in non-Gaussian settings.
Satisfactory approximation of the posterior distribution and associated statistics using MCMC requires evaluation of the simulator at large number of input settings, often in the range of $10^3-10^6$.
Collection of large number of samples become computationally prohibitive  for simulation of a large scale system, imposing key challenge for implementation of the Bayesian framework.
To make the Bayesian framework accessible to large-scale problems, it is necessary to develop computationally efficient uncertainty propagation and calibration techniques.

Marzouk et al. \cite{MarzoukJCP07} have proposed computationally efficient implementation of the Bayesian framework using stochastic spectral methods.
Stochastic spectral projection (SSP) based methods are extensively used for uncertainty propagation as a computationally efficient alternative to Monte Carlo methods with comparable accuracy.
Homogeneous Chaos theory introduced by Wiener \cite{WienerAJM38,Wiener} is an earliest exposition of SSP method, where random variables are represented as an expansion series in orthogonal Hermite polynomials that converges in mean-square sense \cite{CameronAM47}.
Present state of the art in the field of SSP based methods for uncertainty propagation is based on the generalized Polynomial Chaos (gPC) method.
The method has been successfully implemented for solution of stochastic finite element methods \cite{Ghanem91,GhanemPhyD99,Ghanem} and stochastic fluid flow problems \cite{Knio_FDR_2006,Xiu_SJSC_2004}.
Xiu and Karniadakis \cite{XiuJCP03} have extended the method to a set of Askey scheme of orthogonal polynomials.
Subsequently, the method has been applied by various researchers for uncertainty propagation through simulators of systems of engineering importance \cite{Lucor03,Mathelin04,Narayanan04,PoetteJCP09}.
The Bayesian calibration formulation proposed by Marzouk et al. \cite{MarzoukJCP07} uses the gPC method to propagate the prior parametric uncertainty to the simulator prediction.
Resultant gPC expansion of the prediction is used in the Bayes theorem to define the likelihood.
The methodology is further extended by Marzouk and Najm \cite{MarzoukJCP09} for inference of spatially/temporally varying uncertain parameters.

Although the gPC method provides computationally efficient estimation of the uncertainty, computational cost of the implementation grows significantly as the number of stochastic dimensions increases \cite{pierre_12}.
Such a high dimensional uncertainty typically arises for a simulator with a large number of uncertain parameters, and more predominantly in the case of a spatially/temporally varying uncertain parameter with rapidly decaying covariance functions.
The research work presented in this paper addresses the Bayesian framework for calibration of a simulator with high dimensional uncertainty.

Sapsis and Lermusiaux \cite{pierre_09} have proposed dynamically orthogonal field equations (DOFE) method for efficient propagation of high-dimensional uncertainty. The method uses decomposition of the system response into a mean and stochastic dynamical component using a truncated generalized Karhunnen-Loeve expansion.
The stochastic component is spectrally represented in terms of orthogonal eigenfunction basis in spatial dimension, while the respective coefficients define the time varying stochastic dimension.
The Dynamic Orthogonality (DO) condition \cite{pierre_09} is used to derive the closed form evolution equations for the mean, eigenfunction basis and the stochastic coefficients.

However, the dynamic evolution equations of Sapsis and Lermusiaux \cite{pierre_09} does not impose any geometric structure on stochastic dimension, which makes it hard to directly apply the DOFE methodology for Bayesian calibration problems.  This paper proposes a dynamic {\emph bi-orthogonality}-based approach that extends the DOFE method for the Bayesian calibration of a computer simulator with high dimensional uncertainty.
The proposed bi-orthogonal method uses spectral expansion of the stochastic field in the gPC basis, imposing geometric structure on the stochastic dimension. The random coefficients of the truncated generalized Karhunnen-Loeve expansion, obtained using dynamically orthogonal field equations, are projected on the gPC basis using Galerkin projection \cite{Ghanem}.
The resultant field equations are termed here as dynamically bi-orthogonal field equations (DBFE).
The Bayesian calibration approach proposed in this paper uses DBFE to project the prior parametric uncertainty to the system response.
The resultant bi-orthogonal expansion is used in the likelihood during MCMC sampling for Bayesian inference of the uncertain parameters. The proposed DBFE method provides substantial computational speedup over gPC based method for Bayesian calibration. Efficacy of the proposed method is demonstrated for calibration of a 2D transient diffusion equation simulator with uncertain source location and the diffusivity field. Note that a preliminary version of this work was reported in \cite{TagadeIDETC12}, while this article includes (a) extension of the method to take into account model structural uncertainty; (b) substantially refined and expanded theoretical analysis; and (c) additional numerical results demonstrating computational efficiency of the proposed methodology.

The rest of the paper is organized as follows. In section 2, statistical formulation is discussed in detail.
The proposed DBFE-based Bayesian method is presented in section 3.
Section 4 provides numerical results for 2D transient diffusion equation.
Finally, the paper is summarized and concluded in section 5.

\section{Statistical Formulation}
The proposed Bayesian framework is developed for a simulator $T(\boldsymbol{x},t,\boldsymbol{\theta}(\omega))$, where $t\in \mathbb{R}_{\geq 0}$ is time, $\boldsymbol{x}\in \mathcal{X} \subset \mathbb{R}^d$ is a spatial dimension and $\boldsymbol{\theta}(\omega)$ is a set of uncertain parameters that induces uncertainty in the predictions.
Note that the simulator $T(\boldsymbol{x},t,\boldsymbol{\theta}(\omega))$ is defined over a probability space $\left(\Omega, \mathcal{F}, \mathcal{P} \right)$, where $\omega \in \Omega$ is a set of elementary events, $\mathcal{F}$ is associated $\sigma$-algebra and $\mathcal{P}$ is a probability measure defined over $\mathcal{F}$.
In this paper, the proposed method is particularly developed for simulators with a model given by the partial differential equation of type
\begin{equation}
\frac{\partial u(\boldsymbol{x},t;\omega)}{\partial t} = \mathcal{L} \left[u(\boldsymbol{x},t;\omega);\boldsymbol{\theta}(\omega) \right],
\label{spde} \tag{\textbf{SPDE}}
\end{equation}
where $u(\boldsymbol{x},t;\omega)$ is the system response and $\mathcal{L}$ is an arbitrary differential operator.
Equation (\ref{spde}) is known as stochastic partial differential equation (SPDE).
 (\ref{spde}) is initialized using a random field $u(\boldsymbol{x},0;\omega)$, while, the boundary condition is given by
\begin{equation}
\mathcal{B}(\boldsymbol{\beta},t;\omega) = h(\boldsymbol{\beta},t;\omega); ~ ~ ~ \boldsymbol{\beta} \in \partial \mathcal{X}, \omega \in \Omega,
\end{equation}
where $\mathcal{B}$ is a linear differential operator.

The simulator $T(\boldsymbol{x},t,\boldsymbol{\theta}(\omega))$ approximates the physical phenomena within the limits of available knowledge.
Let $\zeta(\boldsymbol{x},t)$ be the `true' but unknown model that perfectly represents the physical phenomena.
Since $T(\boldsymbol{x},t,\boldsymbol{\theta}(\omega))$ is an approximate representation of the physical phenomena, the simulator predictions deviates from $\zeta(\boldsymbol{x},t)$ by $\delta(\boldsymbol{x},t)$, where $\delta(\boldsymbol{x},t)$ is known as the discrepancy function.
The relationship between $\zeta(\boldsymbol{x},t)$ and $T(\boldsymbol{x},t,\boldsymbol{\theta}(\omega))$ is given by \cite{KennedyJRSS01}
\begin{equation}
\zeta(\boldsymbol{x},t) = T(\boldsymbol{x},t,\boldsymbol{\hat{\theta}}(\omega)) + \delta(\boldsymbol{x},t),
\label{relatn_mod_true}
\end{equation}
where $\boldsymbol{\hat{\theta}}(\omega)$ denotes the `true' value of the uncertain parameters.

The proposed Bayesian framework uses experimental observations at finite locations for inference of the uncertain parameters.
At time $t$, let the system be experimentally observed at $M$ spatial locations $\{\boldsymbol{x}_i; i=1,\dots, M\}$.
The measurement at $\boldsymbol{x}_i$, denoted as $y_e(\boldsymbol{x}_i,t)$, is given by
\begin{equation}
y_e(\boldsymbol{x}_i,t) = \zeta(\boldsymbol{x}_i,t) + \epsilon(\boldsymbol{x}_i,t),
\end{equation}
where $\epsilon(\boldsymbol{x}_i,t)$ is the measurement uncertainty.
Let $\boldsymbol{y}_e=\{y_e(\boldsymbol{x}_i,t);~ i=1,...,M\}$ be the set of available experimental observations.
Using $\boldsymbol{y}_e$, the uncertain parameters $\boldsymbol{\theta}$ and the discrepancy function $\delta(\boldsymbol{x},t)$ can be inferred through the Bayes theorem as
\begin{equation}
f\left(\boldsymbol{\hat{\theta}}(\omega),\delta(\boldsymbol{x},t) \mid \boldsymbol{y}_e \right) \propto f\left(\boldsymbol{y}_e\mid T(\boldsymbol{x},t,\boldsymbol{\hat{\theta}}(\omega)),\delta(\boldsymbol{x},t) \right) \times f\left(\boldsymbol{\hat{\theta}}(\omega),\delta(\boldsymbol{x},t) \right),
\label{BayTh}
\end{equation}
where $f(\boldsymbol{\hat{\theta}}(\omega),\delta(\boldsymbol{x},t))$ is the prior, $f(\boldsymbol{y}_e\mid T(\boldsymbol{x},t,\boldsymbol{\hat{\theta}}(\omega)),\delta(\boldsymbol{x},t))$ is the likelihood and $f(\boldsymbol{\hat{\theta}}(\omega),\delta(\boldsymbol{x},t)\mid \boldsymbol{y}_e)$ is the posterior probability distribution.

Uncertainty in the experimental observations, $\epsilon(\boldsymbol{x},t)$, is assumed to be specified using a zero mean Gaussian distribution with covariance matrix
\begin{equation}
\Sigma_e = \sigma^2_e I_M,
\label{exp_unc}
\end{equation}
where  $I$ is the $M\times M$ identity matrix.
In this paper, the proposed Bayesian framework is developed for independent prior uncertainty in $\boldsymbol{\hat{\theta}}(\omega)$ and $\delta(\boldsymbol{x},t)$, with the prior for $\delta(\boldsymbol{x},t)$ given by a zero-mean Gaussian process  with a covariance function of the form
\begin{equation}
\Sigma_\delta(\boldsymbol{x}_1,\boldsymbol{x}_2) = \sigma^2_\delta \exp \left(-\sum^{d}_{i=1} \lambda_i \left(x^i_1 - x^i_2 \right)^2 \right),
\end{equation}
where $\sigma^2_\delta$ is the variance and $\lambda_i$ is the correlation strength of the covariance function, which are treated as uncertain hyper-parameters.  In the present paper, inverse Gamma prior $IG(\alpha_\sigma,\beta_\sigma)$ is used for $\sigma^2_\delta$, while, the Gamma prior $G(\alpha_{\lambda_i},\beta_{\lambda_i})$ is used for $\lambda_i$ \cite{KennedyJRSS01,PauloAS05,O'HaganRESS06}. The uncertainty in $\delta(\boldsymbol{x},t)$ is specified using hierarchical zero-mean Gaussian process prior as
\begin{equation}
f(\delta(\boldsymbol{x},t),\sigma^2_\delta,\lambda_i) \propto \mid \Sigma_\delta \mid^{-\frac{1}{2}} \exp\left( -\frac{1}{2} \boldsymbol{\delta}^T \Sigma^{-1}_\delta \boldsymbol{\delta} \right) \times (\sigma^2_\delta)^{-\alpha_\sigma-1} \exp\left( -\frac{\beta_\sigma}{\sigma^2_\delta} \right) \times \prod^{d}_{i=1} (\lambda_i)^{\alpha_{\lambda_i}-1} \exp\left( -\beta_{\lambda_i} \lambda_i \right)
\label{pr_disc}
\end{equation}
where $\boldsymbol{\delta}=\{\delta(\boldsymbol{x}_i,t);~ i=1,...,N \}$.
Use (\ref{pr_disc}) and (\ref{exp_unc}) in the Bayes theorem (\ref{BayTh}) and marginalize $\delta(\boldsymbol{x},t)$ to obtain
\begin{equation}
f(\boldsymbol{\hat{\theta}}(\omega),\sigma^2_\delta,\lambda_i) \propto \mid \Sigma \mid^{-\frac{1}{2}} \exp\left( -\frac{1}{2} \boldsymbol{\eta}^T \Sigma^{-1} \boldsymbol{\eta} \right) \times (\sigma^2_\delta)^{-\alpha_\sigma-1} \exp\left( -\frac{\beta_\sigma}{\sigma^2_\delta} \right) \times \prod^{d}_{i=1} (\lambda_i)^{\alpha_{\lambda_i}-1} \exp\left( -\beta_{\lambda_i} \lambda_i \right) \times f(\boldsymbol{\hat{\theta}}(\omega)),
\label{BayesFin}
\end{equation}
where $\Sigma=\Sigma_\delta + \Sigma_e$ and $\boldsymbol{\eta}=\{y_e(\boldsymbol{x}_i,t)- T(\boldsymbol{x}_i,t,\boldsymbol{\hat{\theta}}(\omega)); ~ i=1,...,M\}$.

\section{Proposed Methodology}
Equation (\ref{BayesFin}) can be solved by sampling from the posterior distribution using MCMC, which requires evaluation of $T(\boldsymbol{x}_i,t,\boldsymbol{\hat{\theta}}(\omega))$ for each sample, which is computationally prohibitive for large-scale system simulators. The approach proposed in this paper requires single evaluation of (\ref{spde}) using dynamically bi-orthogonal field equations.
The DBFE is used for propagating the prior uncertainty in $\boldsymbol{\hat{\theta}}(\omega)$ to the system response.
The resultant bi-orthogonal expansion of the system response is used in (\ref{BayesFin}) to define the posterior distribution, which is explored using the MCMC.
The proposed method is described in detail in this section.
\subsection{Dynamically Bi-orthogonal Field Equations}
The proposed DBFE method is based on the dynamically orthogonal field equations (DOEF) proposed by Sapsis and Lermusiaux~ \cite{pierre_09}.
Consider a generic Karhunnen-Loeve expansion of $u(\boldsymbol{x},t;\omega)$ truncated at $N$ terms as
\begin{equation}
u(\boldsymbol{x},t;\omega) = \overline{u}(\boldsymbol{x},t) + \sum^{N}_{i=1} Y_i(t;\omega) u_i(\boldsymbol{x},t),
\label{kl_expn_generic}
\end{equation}
where $\overline{u}(\boldsymbol{x},t)$ is the mean, $u_i(\boldsymbol{x},t)$ are the functions that form the complete orthonormal basis on $L^2(\mathcal{X})$, while $Y_i(t;\omega)$ are the zero-mean independent random variables.
Note that throughout this paper the equality sign, $=$, is used to represent the approximate equality, if no confusion is expected. Substituting the expansion (\ref{kl_expn_generic}) in (\ref{spde}) gives
\begin{equation}
\frac{\partial \overline{u}(\boldsymbol{x},t)}{\partial t} + \sum_{i=1}^{N} u_i(\boldsymbol{x},t) \frac{d Y_i(t;\omega)}{d t} + \sum_{i=1}^{N} Y_i(t;\omega) \frac{\partial u_i(\boldsymbol{x},t)}{\partial t} = \mathcal{L}[u(\boldsymbol{x},t;\omega);\boldsymbol{\theta}(\omega)].
\label{gov_eq_doef}
\end{equation}
Note that the quantities $\overline{u}(\boldsymbol{x},t)$, $Y_i(t;\omega)$ and $u_i(\boldsymbol{x},t)$ are dependant on each other through (\ref{kl_expn_generic}).
Thus, varying $\overline{u}(\boldsymbol{x},t)$, $Y_i(t;\omega)$ and $u_i(\boldsymbol{x},t)$ concurrently makes (\ref{gov_eq_doef}) redundant, necessitating imposition of the additional constraint to derive the independent evolution equations for unknown quantities.

Sapsis and Lermusiaux~\cite{pierre_09} proposed imposition of dynamic orthogonality (DO) condition to derive the independent evolution equations for $\overline{u}(\boldsymbol{x},t)$, $Y_i(t;\omega)$ and $u_i(\boldsymbol{x},t)$.
The DO condition constraints the time evolution of $u_i(\boldsymbol{x},t)$ such that
\begin{equation}
\left \langle \frac{\partial u_i(\boldsymbol{x},t)}{\partial t}, u_j(\boldsymbol{x},t) \right \rangle_{\boldsymbol{\boldsymbol{X}}} = 0; ~ ~ ~ \forall i,j=1,...,N,
\label{do_cond}
\end{equation}
where $\left \langle \cdot ,\cdot \right \rangle$ is inner product.
Note that the DO condition ensures that $u_i(\boldsymbol{x},t)$ preserves orthonormality over the time evolution of (\ref{spde}).
\newtheorem{rmkinp}{Remark}
\begin{rmkinp}
In this paper, inner product is defined over spatial and stochastic dimensions.
Inner product over the spatial dimension is defined as
\begin{equation}
\left \langle u(\boldsymbol{x},t;\omega), v(\boldsymbol{x},t;\omega) \right \rangle_{\mathcal{\boldsymbol{X}}} = \int_{\mathcal{\boldsymbol{X}}} u(\boldsymbol{x},t;\omega) v(\boldsymbol{x},t;\omega) d\boldsymbol{x},
\end{equation}
while the inner product over the stochastic dimension is defined as
\begin{equation}
\left \langle u(\boldsymbol{x},t;\omega), v(\boldsymbol{x},t;\omega) \right\rangle_{\Omega}  = \int_{\Omega} u(\boldsymbol{x},t;\omega) v(\boldsymbol{x},t;\omega) d\mathcal{P}(\omega).
\end{equation}
\end{rmkinp}

Using the DO condition, the independent evolution equations for $\overline{u}(\boldsymbol{x},t)$, $u_i(\boldsymbol{x},t)$ and $Y_i(t;\omega)$ are derived as follows \cite{pierre_09}.
\subsubsection{Dynamically Orthogonal Field Equations}
Apply the expectation operator to (\ref{gov_eq_doef}) to obtain the evolution equations for $\overline{u}(\boldsymbol{x},t)$ as
\begin{equation}
\frac{\partial \overline{u}(\boldsymbol{x},t)}{\partial t} = E^{\omega}\left[ \mathcal{L}[u(\boldsymbol{x},t;\omega);\boldsymbol{\theta}(\omega)] \right].
\label{mean_ev}
\end{equation}
Multiply (\ref{gov_eq_doef}) by $Y_j(t;\omega)$ and apply the expectation operator to have
\begin{equation}
\sum^{N}_{i=1} C_{Y_i(t)Y_j(t)} \frac{\partial u_i(\boldsymbol{x},t)}{\partial t} + \sum^{N}_{i=1} C_{\frac{d Y_i(t;\omega)}{dt} Y_j(t;\omega)} u_i(\boldsymbol{x},t) = E^{\omega}\left[\mathcal{L}\left[u(x,t;\omega);\boldsymbol{\theta}(\omega)\right] Y_j(t;\omega)\right],
\label{cov_dyidt}
\end{equation}
where $C_{Y_i(t) Y_j(t)}$ denote the covariance between $Y_i(t;\omega)$ and $Y_j(t;\omega)$.
By multiplying $u_k(\boldsymbol{x},t)$ to (\ref{cov_dyidt}), taking the inner product and applying the expectation operator gives
\begin{equation}
 C_{\frac{d Y_k(t)}{dt} Y_j(t)} = \left\langle E^{\omega}\left[ \mathcal{L}\left(u(x,t;\omega);\omega\right) Y_j(t;\omega)\right], u_k(x,t)\right\rangle_{\mathcal{\boldsymbol{X}}},
\label{cov_dyidtuk}
\end{equation}
which on substitution in (\ref{cov_dyidt}) provides
\begin{equation}
\begin{split}
&\sum^{N}_{i=1} C_{Y_i(t) Y_j(t)} \frac{\partial u_i(\boldsymbol{x},t)}{\partial t} \\& \qquad  =  E^{\omega}\left[\mathcal{L}[u(\boldsymbol{x},t,\omega);\boldsymbol{\theta}(\omega)] Y_j(t;\omega) \right]
 - \sum^{N}_{k=1} \left\langle E^{\omega}\left[ \mathcal{L}[u(\boldsymbol{x},t,\omega);\boldsymbol{\theta}(\omega)] Y_j(t;\omega)\right],u_k(\boldsymbol{x},t) \right\rangle_{\mathcal{\boldsymbol{X}}} u_k(\boldsymbol{x},t).
\end{split} \label{efun_gv}
\end{equation}
Equation (\ref{efun_gv}) can be written in the matrix form
\begin{equation}
{\bf U} = \boldsymbol{\Gamma}^{-1} {\bf D},
\label{efun_gv_mat}
\end{equation}
where $\boldsymbol{\Gamma}$ is the covariance matrix with $(i,j)^{th}$ element $\Sigma_{ij} = C_{Y_i(t) Y_j(t)}$.

To derive the evolution equation for $Y_j(t;\omega)$, multiply both sides of (\ref{gov_eq_doef}) by $u_j(\boldsymbol{x},t)$ and take the inner product to obtain
\begin{equation}
\begin{split}
&\left \langle \frac{\partial \overline{u}(\boldsymbol{x},t)}{\partial t}, u_j(\boldsymbol{x},t) \right \rangle_{\mathcal{\boldsymbol{X}}} + \sum^{N}_{i=1} \left \langle u_i(\boldsymbol{x},t),u_j(\boldsymbol{x},t) \right \rangle_{\mathcal{\boldsymbol{X}}} \frac{d Y_i(t;\omega)}{dt}+ \sum^{N}_{i=1} Y_i(t;\omega) \left \langle \frac{\partial u_i(\boldsymbol{x},t)}{\partial t}, u_j(\boldsymbol{x},t) \right \rangle_{\mathcal{\boldsymbol{X}}}  \\ & \qquad =
\left \langle \mathcal{L}[u(\boldsymbol{x},t;\omega);\boldsymbol{\theta}(\omega)], u_j(\boldsymbol{x},t) \right \rangle_{\mathcal{\boldsymbol{X}}}.
\end{split}\label{dyidt_int_1}
\end{equation}
Note that the third term of left hand side in (\ref{dyidt_int_1}) vanishes completely due to the DO condition (\ref{do_cond}), while, the second term vanishes  for all $i\neq j$ owing to the orthonormality of $u_i(\boldsymbol{x},t)$, thus
\begin{equation}
\frac{d Y_i(t;\omega)}{dt} + \left \langle \frac{\partial \boldsymbol{u}(\boldsymbol{x},t)}{\partial t}, u_i(\boldsymbol{x},t) \right \rangle_{\mathcal{\boldsymbol{X}}} = \left \langle \mathcal{L}[u(\boldsymbol{x},t;\omega);\boldsymbol{\theta}(\omega)], u_i(\boldsymbol{x},t) \right \rangle_{\mathcal{\boldsymbol{X}}}.
\label{dyidt_int_2}
\end{equation}
Note that multiplying (\ref{mean_ev}) by $u_i(\boldsymbol{x},t)$ and taking inner product gives
\begin{equation}
\left \langle \frac{\partial \overline{u}(\boldsymbol{x},t)}{\partial t}, u_i(\boldsymbol{x},t) \right \rangle_{\mathcal{\boldsymbol{X}}} = \left \langle  E^{\omega}\left[ \mathcal{L}[u(\boldsymbol{x},t;\omega);\boldsymbol{\theta}(\omega)] \right], u_i(\boldsymbol{x},t) \right \rangle_{\mathcal{\boldsymbol{X}}}.
\label{dyidt_int_3}
\end{equation}
Using (\ref{dyidt_int_3}) in (\ref{dyidt_int_2}) gives the evolution equation for $Y_i(t;\omega)$ as
\begin{equation}
\frac{d Y_i(t;\omega)}{d t} = \left \langle \mathcal{L}\left[u(\boldsymbol{x},t;\omega);\boldsymbol{\theta}(\omega)\right] -  E^{\omega}\left[\mathcal{L}\left[u(\boldsymbol{x},t;\omega);\boldsymbol{\theta}(\omega) \right]\right], u_i(\boldsymbol{x},t)\right \rangle_{\mathcal{\boldsymbol{X}}}.
\label{yi_ev}
\end{equation}

\subsubsection{Bi-orthogonal Expansion}
Note that the numerical solution of (\ref{spde}) using the DOFE method provide the samples of $u(\boldsymbol{x},t;\omega)$ through the coefficients $Y_i(t;\omega)$, whereas, the Bayesian inference requires analytic form of the probability distribution of the system response, thus, the DOFE method can not directly be used for the Bayesian inference.
In the present paper, a bi-orthogonal expansion approach is proposed to impose the orthogonality based geometric structure on the stochastic dimension.
Consider a gPC expansion of $Y_i(t;\omega)$ truncated at $P$ terms as
\begin{equation}
Y_i(t;\omega) = \sum^{P}_{p=1} Y^i_p(t) \psi_p(\boldsymbol{\xi}(\omega)),
\label{gpc_yi}
\end{equation}
where $\psi_p(\boldsymbol{\xi}(\omega))$ are the orthogonal polynomials from the Askey scheme, while $\boldsymbol{\xi}(\omega)\in L^2(\Xi)$ are the random variables with appropriate probability density function \cite{XiuJCP03}.
Use (\ref{gpc_yi}) in (\ref{kl_expn_generic}) to get
\begin{equation}
u(\boldsymbol{x},t;\omega) = \overline{u}(\boldsymbol{x},t) + \sum^{N}_{i=1} \sum^{P}_{p=1} Y^i_p(t) \psi_p(\boldsymbol{\xi}(\omega)) u_i(\boldsymbol{x},t).
\label{kl_expn_bi}
\end{equation}
Equation (\ref{kl_expn_bi}) is termed here as the {\emph bi-orthogonal} expansion.
Differentiate (\ref{gpc_yi}) with respect to time and use the Galerkin projection to obtain
\begin{equation}
\frac{d Y^i_p(t)}{dt} = \frac{1}{\left\langle \psi^2_q\right\rangle_{\Omega}} \Big\langle \left\langle F\left[u(\boldsymbol{x},t;\omega);\boldsymbol{\theta}(\omega)\right] - E^\omega\left[F\left[u(x,t;\omega);\boldsymbol{\theta}(\omega)\right]\right], u_i(\boldsymbol{x},t) \right\rangle_{\boldsymbol{\boldsymbol{X}}}, \psi_p(\boldsymbol{\xi}(\omega)) \Big\rangle_{\Omega}.
\label{yip_ev}
\end{equation}

Equations (\ref{mean_ev}), (\ref{efun_gv}) and (\ref{yip_ev}) forms dynamically bi-orthogonal field equations (DBFE) that define the dynamic evolution of the mean $\overline{u}(\boldsymbol{x},t)$, the eigenfield $u_i(\boldsymbol{x},t)$ and the associated coefficients $Y^i_p(t)$.
The resultant bi-orthogonal expansion (\ref{kl_expn_bi}) approximates the system response $u(\boldsymbol{x},t;\omega)$ to an arbitrary accuracy depending on the number of eigenfunctions used, $N$, and the number of expansion coefficients for each eigenfunction, $P$.

\subsubsection{Boundary Conditions}
To define boundary conditions for DBFE, consider a generic Karhunnen-Loeve
expansion of $h(\boldsymbol{\beta},t;\omega)$
\begin{equation}
h(\boldsymbol{\beta},t;\omega) = \overline{h}(\boldsymbol{\beta},t) + \sum^{N}_{i=1} Y_i(t;\omega) u_i(\boldsymbol{\beta},t).
\label{kl_bound}
\end{equation}
Applying the expectation operator to (\ref{kl_bound}), boundary condition for the mean is given by
\begin{equation}
\mathcal{B}(\overline{u}(\boldsymbol{x},t)) = \overline{h}(\boldsymbol{\beta},t).
\end{equation}
By multiplying $Y_j(t;\omega)$ to (\ref{kl_bound}) and applying the expectation operator to obtain boundary condition for $u_i(\boldsymbol{x},t)$
\begin{equation}
\mathcal{B}(u_i(\boldsymbol{\beta},t)) = \sum^{N}_{j=1} C^{-1}_{Y_i(t) Y_j(t)} E^{\omega}\left[h(\boldsymbol{\beta},t;\omega) Y_j(t;\omega)\right].
\end{equation}

\subsection{Bayesian Inference}
Without loss of generality, the proposed method is described here for a spatially varying uncertain parameter with prior given by a scalar stochastic process $v(\boldsymbol{x};\omega)$,
i.e. $\boldsymbol{\theta}(\omega)=\{v(\boldsymbol{x};\omega)\}$ .
Use a KL expansion of $v(\boldsymbol{x};\omega)$ as
\begin{equation}
v(\boldsymbol{x};\omega) = \overline{v}(\boldsymbol{x}) + \sum^{N}_{i=1} \sqrt{\lambda_i} v_i(\boldsymbol{x}) \chi_i,
\label{kl_theta}
\end{equation}
where $\chi_i$ are independent identically distributed zero-mean random variables, while, $\lambda_i$ and $v_i(\boldsymbol{x})$ are the eigenvalues and eigenfunctions of the covariance function of $v(\boldsymbol{x};\omega)$.
For the covariance function $C_v(\boldsymbol{x}_1,\boldsymbol{x}_2)$, $\lambda_i$ and $v_i(\boldsymbol{x})$ are solution of the eigenvalue problem
\begin{equation}
\int_{\mathcal{X}} C_v(\boldsymbol{x}_1,\boldsymbol{x}_2) \nu_i(\boldsymbol{x}_1) d\boldsymbol{x}_1 = \lambda_i v_i(\boldsymbol{x}_2).
\end{equation}
For a Gaussian process prior, $\chi_i$ are standard normal random variables, whereas, for a generic stochastic process prior, $\chi_i$ are given by
\begin{equation}
\chi_i = \frac{1}{\sqrt{\lambda_i}} \int_{\mathcal{X}} \left(v(\boldsymbol{x};\omega) - \overline{v}(\boldsymbol{x}) \right) v_i(\boldsymbol{x}) d\boldsymbol{x}.
\end{equation}

Use the gPC expansion of $\chi_i$
\begin{equation}
\chi_i = \sum^{P}_{p=1} \hat{\chi}^i_p \psi_p(\boldsymbol{\xi}(\omega)),
\label{chi_gpc}
\end{equation}
where $\hat{\chi}^i_p$ are the gPC expansion coefficients, in (\ref{kl_theta}) to get the bi-orthogonal expansion of $v(\boldsymbol{x};\omega)$ as
\begin{equation}
v(\boldsymbol{x};\omega) = \overline{v}(\boldsymbol{x}) + \sum^{N}_{i=1} \sum^{P}_{p=1}\sqrt{\lambda_i} v_i(\boldsymbol{x}) \hat{\chi}^i_p \psi_p(\boldsymbol{\xi}(\omega)).
\label{theta_bi_expn}
\end{equation}
The expansion (\ref{theta_bi_expn}) is used in $\mathcal{L}\left[u(\boldsymbol{x},t;\omega);\boldsymbol{\theta}(\omega)\right]$ to define the RHS of the DBFE governing equations ((\ref{mean_ev}), (\ref{efun_gv}) and (\ref{yip_ev})).
The numerical solution of the resultant DBFE governing equations give the bi-orthogonal expansion (\ref{kl_expn_bi}) of the system response $u(\boldsymbol{x},t;\omega)$.

\newtheorem{rmkini}[rmkinp]{Remark}
\begin{rmkini}
The numerical solution is initiated with the initial condition for the mean $\overline{u}(\boldsymbol{x},t)$ given by
\begin{equation}
\overline{u}(\boldsymbol{x},t) = E^\omega\left[F[u(\boldsymbol{x},0;\omega);\boldsymbol{\theta}(\omega)]\right],
\end{equation}
while, the initial conditions for the eigenfield are given by
\begin{equation}
u_i(\boldsymbol{x},t) = v_i(\boldsymbol{x}).
\end{equation}
Since the stochasticity in (\ref{spde}) emanates due to the uncertainty in $\nu(\boldsymbol{x};\omega)$, (\ref{spde}) is initialized with a deterministic initial condition. Thus, the initial condition for the expansion coefficients $Y^i_p(t)$ is given by
\begin{equation}
Y^i_p(0) = 0; ~ ~ ~ \forall i=1,..,N; ~  p=1,...,P.
\end{equation}
\end{rmkini}

The bi-orthogonal expansion of $u(\boldsymbol{x},t;\omega)$ is used in (\ref{BayesFin}) to define the likelihood
\begin{equation}
f(\boldsymbol{y}_e \mid \boldsymbol{\xi},\sigma^2_\delta,\lambda_i) \propto \mid \Sigma \mid^{-\frac{1}{2}} \exp \left(-\frac{1}{2} \boldsymbol{\eta}^T \Sigma^{-1} \boldsymbol{\eta} \right),
\label{likelihood_xi}
\end{equation}
where
\begin{equation}
\boldsymbol{\eta} = \left\{y_e(\boldsymbol{x}_k,t) - \left(\overline{u}(\boldsymbol{x}_k,t) + \sum^{N}_{i=1} \sum^{P}_{p=1} Y^i_p(t) u_i(\boldsymbol{x}_k,t) \psi_p(\boldsymbol{\xi}(\omega)) \right); ~ ~ ~ k=1,...,M \right\}.
\label{mu_def_xi}
\end{equation}
Note that conditional on the hyper-parameters of the discrepancy function, $\sigma^2_\delta$ and $\lambda_i$, $\boldsymbol{\xi}$ are the only uncertain parameters in (\ref{likelihood_xi}).
Thus, the Bayesian calibration problem is reformulated in the space $L^2(\Xi)$ as the inference of $\boldsymbol{\xi}$.
In the present paper, the proposed method is demonstrated for Hermite polynomials as gPC basis, where $\boldsymbol{\xi}$ are the independent identically distributed standard normal random variables, thus, the prior for $\boldsymbol{\xi}$ is given by
\begin{equation}
f(\boldsymbol{\xi}) \propto \prod^{N_z}_{k=1} \exp\left(-\frac{\xi^2_k}{2} \right),
\label{prior_xi}
\end{equation}
where $N_z$ is the dimension of stochasticity.
Using (\ref{prior_xi}) and (\ref{likelihood_xi}) in (\ref{BayesFin}), the proposed formulation for the Bayesian inference is
\begin{align}
f(\boldsymbol{\xi},\sigma^2_\delta,\lambda_i) & \propto \mid \Sigma \mid^{-\frac{1}{2}} \exp\left( -\frac{1}{2} \boldsymbol{\eta}^T \Sigma^{-1} \boldsymbol{\eta} \right) \times (\sigma^2_\delta)^{-\alpha_\sigma-1} \exp\left( -\frac{\beta_\sigma}{\sigma^2_\delta} \right) \times \prod^{d}_{i=1} (\lambda_i)^{\alpha_{\lambda_i}-1} \exp\left( -\beta_{\lambda_i} \lambda_i \right) \nonumber \\
~ & \times \prod^{N_z}_{k=1}  \exp\left(-\frac{\xi^2_k}{2} \right).
\label{BayesNewForm}
\end{align}
Note that (\ref{BayesNewForm}) does not involve the solution of $T(\boldsymbol{x},t,\boldsymbol{\theta}(\omega))$, thus, the posterior distribution can can be explored efficiently using MCMC.
In the present paper, Metropolis-Hastings algorithm \cite{MetropolisJCP53,HastingsBio70} is used to sample from the posterior distribution.

\section{Numerical Example: 2D Transient Diffusion Equation}
\begin{figure}[t!]
  \begin{center}
  \begin {tabular}{l l l}
  \subfigure [spatial variation of diffusivity]{\includegraphics[width=2in, height=1.85in] {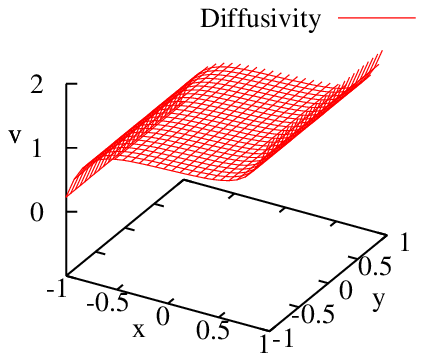}} &
  \subfigure [u-field at $t=0.05s$ ]{\includegraphics[width=2in, height=1.85in] {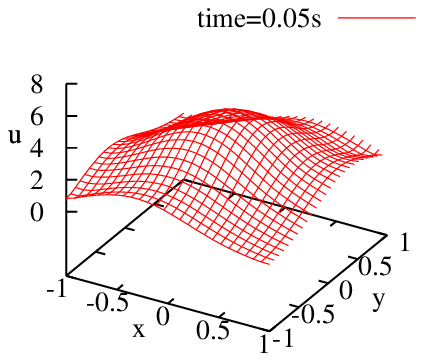}} &
  \subfigure [u-field at $t=0.1s$ ]{\includegraphics[width=2in, height=1.85in] {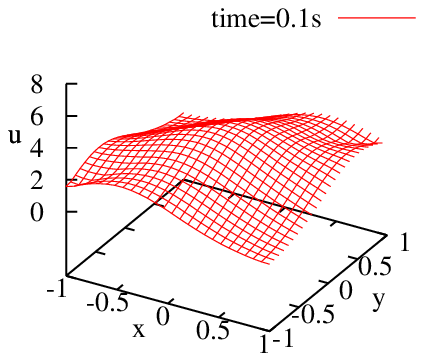}} \\
   \end{tabular}
  \end{center}
  \vspace*{-.2in}
 \caption{Solution of two-dimensional transient diffusion equation}
\label{det_soln}
\end{figure}

The efficacy and efficiency of the proposed method is investigated for calibration of a two-dimensional transient diffusion simulator with uncertain source location and diffusivity field.
Present paper considers a stochastic transient diffusion equation defined over a two-dimensional domain $\mathcal{X}=\left[-1, 1 \right]\times \left[-1, 1 \right]$ with adiabatic boundary conditions as
\begin{align}
\frac{\partial u(\boldsymbol{x},t;\omega)}{\partial t} & = \nabla \left(\nu(\boldsymbol{x};\omega) \nabla u(\boldsymbol{x},t;\omega) \right) + \sum^{N_s}_{l=1} \frac{s_l}{2 \pi \sigma^2} \exp\left(- \frac{\left(\boldsymbol{z}_l - \boldsymbol{x} \right)^2}{2 \sigma^2_l} \right) \\
\nabla u(\boldsymbol{x},t;\omega) \cdot \hat{n} & = 0 \\
u(\boldsymbol{x},t;\omega) & = 0,
\label{twod_trans_diff}
\end{align}
where $\nu(\boldsymbol{x};\omega)$ is spatially varying diffusivity field, while total $N_s$ source term are active at locations $\boldsymbol{z}_l$ with source strength $S_l$.
The diffusivity field, $\nu(\boldsymbol{x};\omega)$, and the location of the source, $\boldsymbol{z}$, are assumed to be uncertain.
Efficacy of the proposed method is demonstrated using the `hypothetical test bed' data, which is defined using the numerical solution of (\ref{twod_trans_diff}) for completely known source location and the diffusivity field.
In the present paper, the proposed method is demonstrated for a single source located at $(0.2,-0.2)$, which is active during time $\left[0,0.01\right]$.
The spatial variation of the diffusivity is assumed to take the form
\begin{equation}
\nu(\boldsymbol{x}) = 0.05(\nu_0 + 10.0 + 0.25 x + 0.65 y + x^3 + y^3),
\label{diff_mean}
\end{equation}
where $\nu_0$ is a user defined constant.
Figure \ref{det_soln}(a) shows spatial variation of the diffusivity.
The deterministic numerical solution is obtained using a second order central difference scheme in spatial dimension with uniform grid spacing $h = 0.02$,  while the explicit fourth order Runge-Kutta scheme is used for time integration with the time step $\Delta t = 0.0001$.
Figure \ref{det_soln}(b) shows the numerical solution at $t=0.05s$, while the solution at $t=0.1s$ is shown in Figure \ref{det_soln}(c).
Note that the source has peak strength at $\boldsymbol{z}_l$ and reduces exponentially with the distance, resulting in the peak value of $u$ at the source location and the subsequent diffusion with time to other locations.
Upon removal of the source, diffusion of $u$ is non-uniform owing to the non-linear diffusivity.
\subsection{DBFE Formulation}
For notational convenience, define
\begin{equation}
S(\boldsymbol{x};\omega) = \frac{s}{2 \pi \sigma^2} \exp\left(- \frac{\left(\boldsymbol{z} - \boldsymbol{x} \right)^2}{2 \sigma^2} \right),
\end{equation}
which is uncertain owing to the uncertainty in the source location $\boldsymbol{z}$.
The prior uncertainty in $\boldsymbol{z}$ is expanded in a gPC basis, while the Galerkin projection is used to obtain the resultant gPC coefficients, $\hat{S}(\boldsymbol{x})$, of
\begin{equation}
S(\boldsymbol{x};\omega) = \sum^{P}_{p=1} \hat{S}(\boldsymbol{x}) \psi_p(\omega).
\label{gpc_s}
\end{equation}
The prior uncertainty in $\nu(\boldsymbol{x};\omega)$ is represented using a Gaussian process, which is spectrally represented using the bi-orthogonal expansion as
\begin{equation}
\nu(\boldsymbol{x};\omega) = \overline{\nu}(\boldsymbol{x};\omega) + \sum^{N}_{i=1} \sum^{P}_{p=1} V^i_p \nu_i(\boldsymbol{x}) \psi_p(\omega),
\label{bi_nu}
\end{equation}
where $\overline{\nu}(\boldsymbol{x};\omega)$ is the mean, $\nu_i(\boldsymbol{x})$ are the eigenfunctions of the covariance function of $\nu(\boldsymbol{x};\omega)$ and $V^i_p$ are the respective expansion coefficients.
Use (\ref{gpc_s}) and (\ref{bi_nu}) in (\ref{twod_trans_diff}) to obtain the differential operator in (\ref{spde}) as
\begin{align}
\begin{aligned}
\mathcal{L}\left[u(\boldsymbol{x},t;\omega);\boldsymbol{\theta}(\omega)\right] & = \nabla[\overline{\nu}(\boldsymbol{x}) \nabla \overline{u}(\boldsymbol{x},t) + \overline{\nu}(\boldsymbol{x}) \sum^{N}_{i=1} \sum^{P}_{p=1} Y^i_p(t) \psi_p(\boldsymbol{\xi}(\omega)) \nabla u_i(\boldsymbol{x},t) \\
~ & + \sum^{N}_{i=1} \sum^{N}_{j=1} \sum^{P}_{p=1} \sum^{P}_{q=1} V^i_p Y^j_q(t) \nu_i(\boldsymbol{x}) \psi_p(\boldsymbol{\xi}(\omega)) \psi_q(\boldsymbol{\xi}(\omega)) \nabla u_j(\boldsymbol{x},t) \\
~ & + \sum^{N}_{i=1} \sum^{P}_{p=1} V^i_p \nu_i(\boldsymbol{x}) \psi_p(\boldsymbol{\xi}(\omega)) \nabla \overline{u}(\boldsymbol{x},t)] + \sum^{P}_{p=1} \hat{S}(\boldsymbol{x}) \psi_p(\boldsymbol{\xi}(\omega)).
\end{aligned}
\label{diff_oper_transdiff}
\end{align}
Use (\ref{diff_oper_transdiff}) in (\ref{mean_ev}), (\ref{efun_gv}) and (\ref{yip_ev}) to obtain the DBFE governing equations for the two-dimensional transient-diffusion equation.

\subsection{Solution of Forward Problem}
The prior uncertainty in the source location is specified using independent Gaussian distribution for $x$ and $y$ co-ordinate with mean $0$ and standard deviation $0.3$, while, the prior for diffusivity $\nu(\boldsymbol{x};\omega)$ is specified using a Gaussian process with mean
\begin{equation}
\nu(\boldsymbol{x}) = 0.05(\nu_0 + 10.0 + 0.25 x + 0.65 y)
\label{mean_diff_unc}
\end{equation}
and the squared exponential covariance function
\begin{equation}
C(\boldsymbol{x}_1,\boldsymbol{x}_2) = \sigma^2_\nu \exp \left(-\lambda_1 (x_1 - x_2)^2 - \lambda_2 (y_1 - y_2)^2 \right),
\label{cov_diff_unc}
\end{equation}
where $\sigma^2_\nu$ is the variance of the Gaussian process and $\lambda_i$ is the correlation length.

Efficacy and the computational efficiency of the proposed Bayesian inference depends on the ability of the DBFE method to accurately solve the forward forward problem.
In the present subsection, accuracy and the computational cost for the numerical implementation of DBFE is compared against the Monte Carlo and the generalized Polynomial Chaos method (see \cite{MarzoukJCP07} for the gPC formulation of (\ref{twod_trans_diff})).

\begin{figure}[t]
  \begin{center}
  \begin {tabular}{l l}
  \subfigure [CPU time]{\includegraphics[width=3in,height=2.85in] {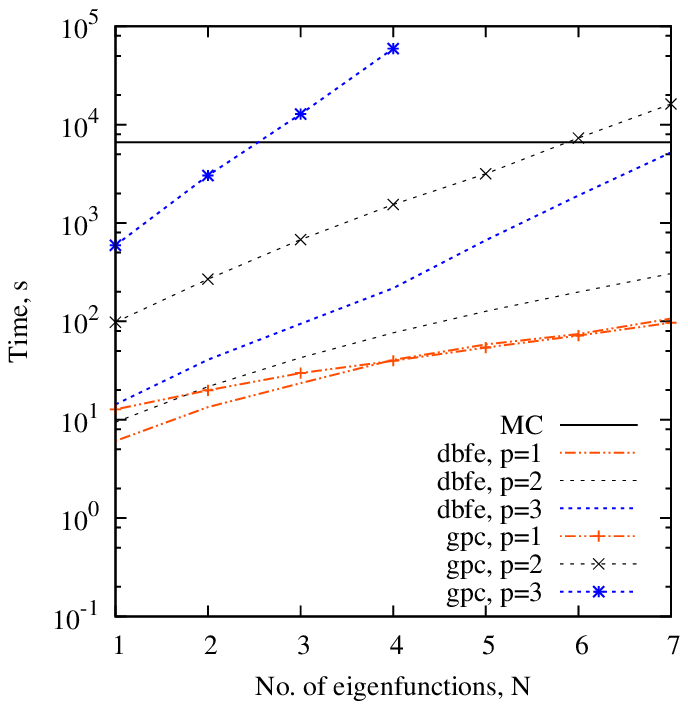}} &
  \subfigure [$L_1$-error]{\includegraphics[width=3in,height=2.85in] {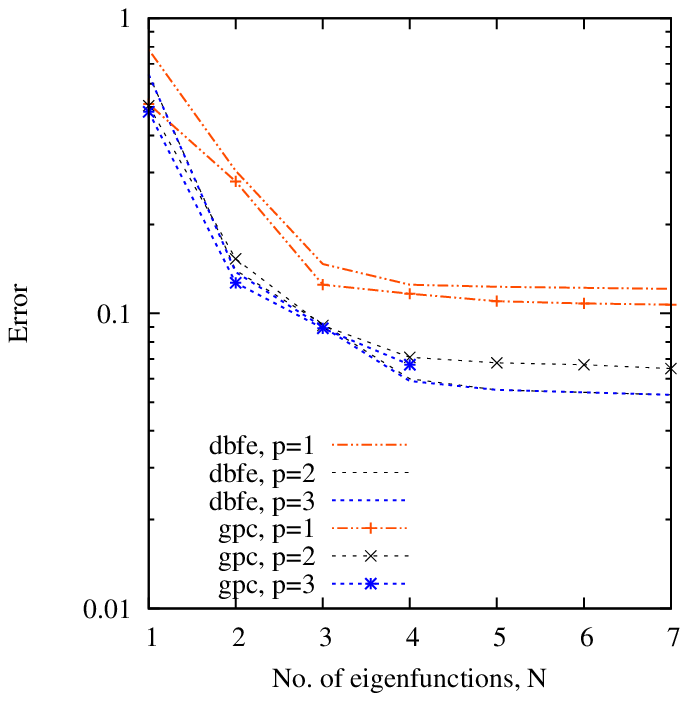}}
   \end{tabular}
  \end{center}
    \vspace*{-.2in}
 \caption{Comparison of accuracy and computational efficiency of DBFE with gPC and Monte Carlo method}
\label{comp_dbfe_gpc}
\end{figure}

Figure \ref{comp_dbfe_gpc} shows the accuracy and computational efficiency of the DBFE and the gPC method for different values of number of eigenfunctions used, $N$, and the order of the polynomial chaos basis, $p$.
The accuracy is compared using the Monte Carlo method with 10000 samples, which are collected at the computational cost of 6616.17 seconds.
The computational cost for solution of the forward problem increases with increase in $N$ and $p$ for both the DBFE and gPC method.
Note that the stochastic dimension for the present problem is $N+2$ ($N$ dimensions representing the truncated KL expansion, while $2$ dimensions representing uncertainty in the source location), for which the number of polynomial chaos terms is given by
\begin{equation}
P = \frac{(N+2+p)!}{(N+2)! p!} + 1.
\label{poly_chaos_order}
\end{equation}
Since implementation of the gPC method requires numerical solution of $P$ PDEs (see \cite{MarzoukJCP09} for details), whereas, the DBFE method involves numerical solution of $(N+1)$ PDEs and $N\times P$ ODEs, the increase in computational cost with $p$ is significantly higher for the gPC method as compared with the DBFE method.
The computational cost of the gPC method is comparable to the Monte Carlo method for $N=6$ and second order polynomial chaos basis, while the computational cost is higher than the Monte Carlo method for third order polynomial chaos basis with $N\geq 3$, rendering the gPC method computationally intractable.
The DBFE method is numerically implemented at a computational cost comparable to the gPC method for $p=1$, while, the computational cost for the DBFE method for $p\geq 2$ is considerably lower than the gPC method.
The $L_1$-error in variance is shown in Figure \ref{comp_dbfe_gpc}(b).
The error is comparable for both the DBFE and the gPC method, which decrease with $N$ and reaches the limiting value for $N\geq 4$, though the limiting value is higher for $p=1$.
Note that the transient diffusion equation involves multiplication of the diffusivity $\nu$ with $\nabla u$, thus, appropriate spectral representation requires use of the second order polynomial chaos basis.
From the results, it may be concluded that (\ref{twod_trans_diff}) can be numerically solved using the DBFE method at significantly lower computational cost than the gPC method with the comparable accuracy.

\subsection{Solution of Source Inversion Problem}
The proposed method is used for inference of the source location and the diffusivity.
Prior uncertainty in the source location is given by independent Gaussian processes in $x$ and $y$ directions, with $\mathcal{N}(0.0,0.3)$ prior.
Prior uncertainty in the diffusivity is specified using the Gaussian process with the mean (\ref{mean_diff_unc})
and the covariance function (\ref{cov_diff_unc})
with $\sigma^2=0.3$ and $\lambda = 1.5$.
The prior uncertainty is propagated to the system response using the DBFE method with $N=5$ and $p=2$.
The deterministic solution of the 2D transient diffusion equation at time $t=0.02$ seconds with the source removed at $t=0.01$ seconds is used as experimental observations.
The source is assumed to be located at $[0.2,-0.2]$.
Total 25 uniformly spaced data points are used for the Bayesian inference.
1\% experimental uncertainty is assumed in each data point.
The model structure uncertainty is defined by specifying the prior probability distribution for $\sigma^2_\delta$ and $\lambda_\delta$.
Inverse Gamma distribution $IG(6.0,2.0)$ is used for $\sigma^2_\delta$, while, the prior for $\lambda_\delta$ is given by the Gamma distribution $G(6.0,2.0)$.
Posterior distribution is explored by collected 10000 samples using MCMC, after burnout period of 1000 samples.

\begin{figure}[t]
\begin{center}
\includegraphics[width=4in,height=3.85in]{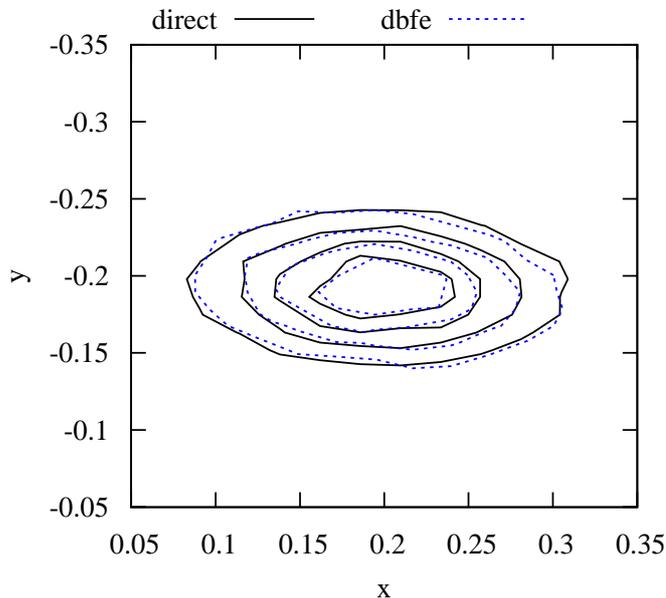}
\end{center}
\vspace*{-.5in}
\caption{Posterior Probability Contours for the Source Location }
\label{cont_pdf_xy}
\end{figure}

Figure \ref{cont_pdf_xy} shows the posterior probability density contours for the source location.
The contour is shown for the posterior density obtained using the proposed method (dashed line) and the direct MCMC sampling from the posterior distribution (solid lines).
The source location is predicted accurately, while, the posterior density obtained using the proposed method agrees closely with the direct MCMC sampling, demonstrating the efficacy of the DBFE based Bayesian inference.

\begin{figure}[t]
  \begin{center}
  \begin {tabular}{l l}
  \subfigure [$L_1$-error in Posterior Variance for DBFE and Direct MCMC Sampling]{\includegraphics[width=3in,height=2.85in]{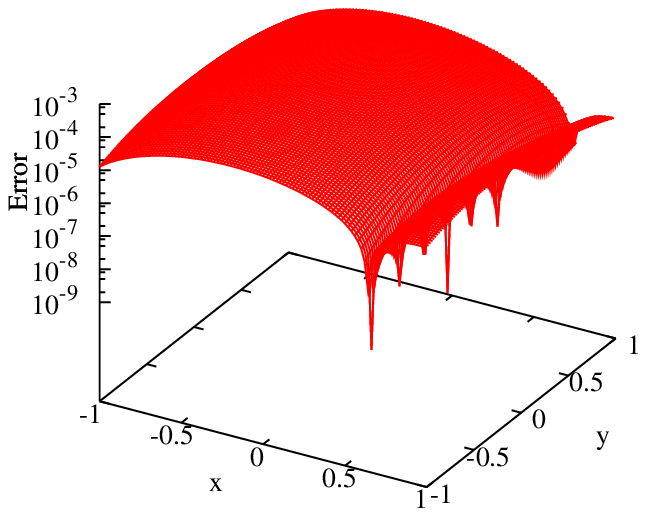}} &
  \subfigure [$L_1$-error in Posterior Mean for DBFE]{\includegraphics[width=3in,height=2.85in]{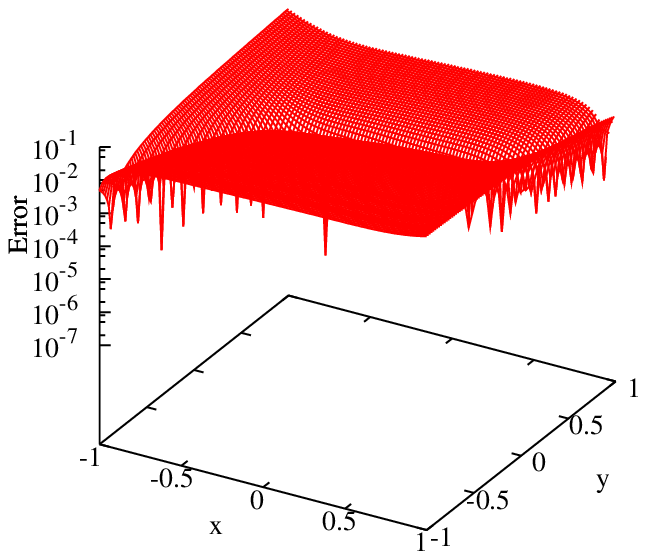}}
   \end{tabular}
  \end{center}
    \vspace*{-.2in}
 \caption{$L_1$-error in posterior variance and mean for diffusivity}
\label{post_mean_v}
\end{figure}

Figure \ref{post_mean_v}(a) shows the $L_1$-error in posterior variance of the diffusivity between the DBFE the direct MCMC sampling.
The maximum error is of the order of $10^{-3}$, indicating the close agreement in variance for the posterior probability of the diffusivity obtained using the DBFE and the direct MCMC sampling methods.
Figure \ref{post_mean_v}(b) shows the $L_1$-error for posterior mean of the diffusivity for the DBFE method obtained against the `true' diffusivity.
Note that the error reduces non-uniformly, indicating the effect of the location of the experimental observations on the proposed Bayesian inference.

\begin{figure}[ht]
  \begin{center}
  \begin {tabular}{l l}
  \subfigure [Probability distribution for $\sigma^2_\delta$]{\includegraphics[width=3in,height=2.85in] {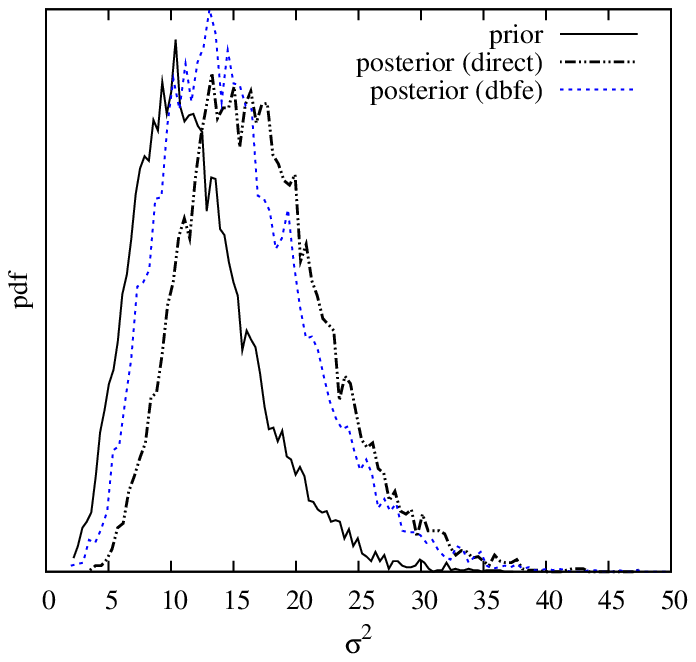}} &
  \subfigure [Probability distribution for $\lambda_\delta$]{\includegraphics[width=3in,height=2.85in] {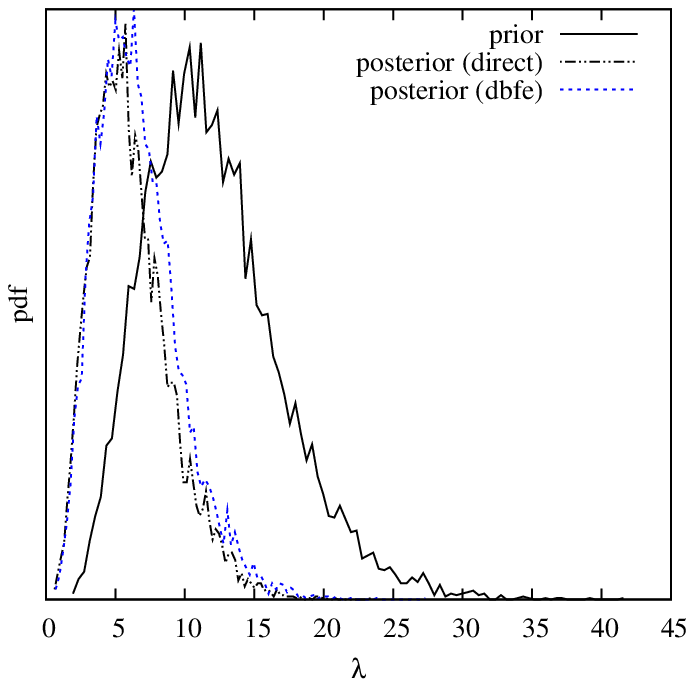}}
   \end{tabular}
  \end{center}
    \vspace*{-.2in}
\caption{Comparison of probability distribution of $\sigma^2_\delta$ and $\lambda_\delta$
 }
\label{post_mean_del}
\end{figure}

Figure \ref{post_mean_del} shows the comparison of the posterior probability distribution of $\sigma^2_\delta$ and $\lambda_\delta$ obtained using the direct MCMC sampling and the proposed method.
The posterior distribution for $\lambda_\delta$ obtained using the proposed method matches closely with the direct sampling, however, the match is comparatively poor for the posterior distribution of $\sigma^2_\delta$.
Note that the bi-orthogonal expansion obtained using the DBFE method acts as an emulator of the 2D transient diffusion equation, which is used in the Bayesian inference as against the simulator in the direct MCMC sampling.
Thus, any remnant error in the bi-orthogonal expansion is considered as the uncertainty in the model structure, resulting in the difference in the posterior probability distribution for $\sigma^2_\delta$.
The posterior probability for $\lambda_\delta$ has moved towards the left for both the cases, indicating increased correlation for the model structure uncertainty, while, the posterior distribution for $\sigma^2_\delta$ moves towards right, indicating higher posterior confidence on the simulator.

\begin{figure}[ht]
\begin{center}
\includegraphics[width=3in,height=2.85in]{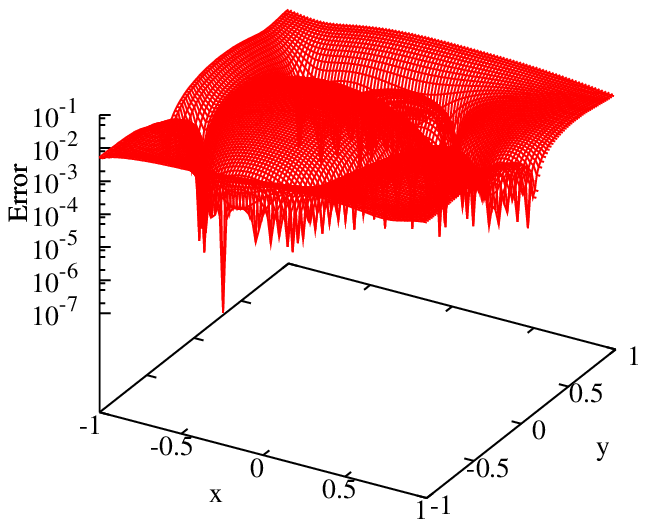}
\end{center}
\vspace*{-.35in}
\caption{$L_1$-error in posterior mean for $u$}
\label{post_mean_u}
\end{figure}

Figure \ref{post_mean_u} shows the $L_1$-error in the posterior mean for the system response $u$, which is defined against the true spatial distribution of $u$.
The posterior mean of the system response is calculated by substituting the mean of $\boldsymbol{\xi}$ in the bi-orthogonal expansion.
The maximum $L_1$-error is of the order of $10^{-1}$, which is located in the boundary region where experimental data is not provided for the Bayesian inference.
In the region where experimental observations are available, the error is significantly low with the minimum value of the order of $10^{-7}$.


\section{Concluding Remarks}
The paper has presented a dynamic bi-orthogonality based approach for computationally efficient implementation of the Bayesian inference. 
The proposed method can be applied for calibration of a simulator represented using partial differential equation with high dimensional uncertainty. 
Though the method requires reformulation of the governing equations, existing schemes can be extended in a straightforward manner for numerical solution of the DBFE.    

Efficacy of the proposed method is demonstrated for calibration of a two-dimensional transient diffusion equation with uncertain source location and the diffusivity. 
Computational cost of the proposed method for uncertainty propagation is compared against the gPC and the Monte Carlo method. 
Note that for low dimensional uncertainty, computational cost of the gPC method is comparable to the DBFE, however, as dimensionality of the uncertainty increases, the DBFE method provide the solution of the SPDE at a significantly less computational cost than the gPC method with comparable accuracy. 
Accuracy of the proposed method to infer the uncertain parameters is compared against the direct MCMC sampling. 
The method provide accurate inference of the source location with the marginal posterior distribution matching closely with the MCMC sampling. 
The method is found to accurately infer the posterior distribution of the spatially/temporally varying parameters. 
In the present paper, the proposed method is demonstrated in the Gaussian context. 
In the future, efficacy of the proposed method will be demonstrated for a more generic non-Gaussian non-stationary setting.


\bibliographystyle{plain}
\bibliography{References_thesis, References}


\end{document}